\documentclass[journal=jacsat,manuscript=article,layout=singlecolumn]{achemso}

\usepackage{chemformula} 
\usepackage[T1]{fontenc} 

\usepackage{amsfonts,mathrsfs,amsmath,amsthm,amssymb,bbold}
\usepackage{epsfig}
\usepackage{bm,mathrsfs}
\usepackage{lineno,hyperref}
\usepackage{color}
\usepackage{xcolor,colortbl} 
\usepackage{csquotes}
\usepackage{soul}
\usepackage{afterpage}
\usepackage{capt-of}
\usepackage{makecell}
\usepackage{graphicx}
\setcellgapes{4pt}

\usepackage{ulem}

\newcolumntype{P}[1]{>{\centering\arraybackslash}p{#1}}

\usepackage{amsmath} 
\newcommand{\angstrom}{\text{\normalfont\AA}}

\clearpage

\author{Hung Ba Tran}
\affiliation{Advanced Institute for Materials Research (WPI-AIMR), Tohoku University, Sendai, Miyagi 980-8577, Japan}
\email{tran.ba.hung.a6@tohoku.ac.jp}

\author{Toyoto Sato} 
\affiliation{Institute for Materials Research (IMR), Tohoku University, Sendai, Miyagi 980-8577, Japan}

\author{Ryuhei Sato} 
\affiliation{Department of Materials Engineering, The University of Tokyo, Tokyo 113-8656, Japan}

\author{Hiroyuki Saitoh} 
\affiliation{National Institutes for Quantum Science and Technology, Sayo-gun, Hyogo 679-5148, Japan}

\author{Shin-ichi Orimo} 
\affiliation{Advanced Institute for Materials Research (WPI-AIMR), Tohoku University, Sendai, Miyagi 980-8577, Japan}
\alsoaffiliation{Institute for Materials Research (IMR), Tohoku University, Sendai, Miyagi 980-8577, Japan}
\email{shin-ichi.orimo.a6@tohoku.ac.jp}

\author{Hao Li} 
\affiliation{Advanced Institute for Materials Research (WPI-AIMR), Tohoku University, Sendai, Miyagi 980-8577, Japan}
\email{li.hao.b8@tohoku.ac.jp}

\title{Tuning Stability of AB$_{3}$-Type Alloys by Suppressing Magnetism}

\date{\today}

\begin{document}

\abstract{
Hydrogen is a promising clean energy carrier, yet effective and reversible storage remains a major challenge. AB$_{3}$-type intermetallic alloys have emerged as promising candidates for solid-state hydrogen storage owing to their intermediate thermodynamic stability and rapid hydrogen uptake. However, optimizing both stability and gravimetric density has been hindered by competing thermodynamic and magnetic effects. Here, we present a systematic analysis of AB$_{3}$ compounds (A = Ca, Y, Mg; B = Co, Ni) and their ternary alloys Ca$_{x}$Y$_{y}$Mg$_{1-x-y}$B$_{3}$ (B = Co, Ni), combining first-principles calculations with Monte Carlo simulations. Strikingly, we uncover a direct correlation between formation energy and total magnetic moment that dictates alloy stability, thereby explaining the observed trade-off in hydrogen storage optimization. In Co-based systems with large lattice volumes, such as Ca-rich and Y-rich compositions, formation energy increases monotonically with magnetization, exhibiting a near one-to-one correlation and establishing magnetism as the dominant factor governing alloy stability. While Mg-rich compositions achieve high gravimetric densities, strong magnetism destabilizes the system, necessitating Y substitution to suppress magnetic moments and stabilize a low-spin state--thereby limiting optimization when heavy Y is introduced.  In contrast, replacing Co with Ni dramatically weakens the magnetism: YNi$_{3}$ is nonmagnetic, while CaNi$_{3}$ and MgNi$_{3}$ display only weak spin polarization, enabling thermodynamic stability across the entire compositional range. Notably, the experimentally known CaMg$_{2}$Ni$_{9}$ combines high theoretical capacity ($\sim$3.32 wt\%) with good reversibility, and our calculations further identify the Mg-rich Ni-based region as an unexplored compositional domain that couples negative formation energies with the highest gravimetric densities (up to $\sim$3.40 wt\%). These findings establish magnetism as a fundamental thermodynamic variable controlling alloy stability and reveal magnetic suppression via transition-metal substitution as the key to overcoming the stability--capacity trade-off in AB$_{3}$-type hydrogen storage materials.}

\section{Introduction}

Hydrogen represents a clean and sustainable energy carrier for automobiles and power generation, supporting the global transition toward a zero-emission society\cite{Staffell2019EES, Pollet2024IJHE, Ahmad2024ACSSRM}. Green hydrogen enables efficient storage of renewable energy from solar and wind sources, which can then be utilized in fuel cell systems\cite{Staffell2019EES, Oliveira2021COCE, Eriksson2017AE}. However, effective hydrogen storage remains one of the most critical challenges in establishing a practical hydrogen-based infrastructure\cite{Allendorf2022NC, Felderhoff2007PCCP}. At ambient conditions, hydrogen gas exhibits extremely low volumetric energy density\cite{Allendorf2022NC}. High-pressure storage increases density but requires heavy, specialized tanks that reduce gravimetric efficiency\cite{Felderhoff2007PCCP}. This limitation parallels those of electric vehicles, where heavy batteries constrain driving range\cite{Deng2020J}. Meanwhile, liquid hydrogen demands cryogenic temperatures (20 K), imposing large energy costs for cooling\cite{Allendorf2022NC, Felderhoff2007PCCP, Zhang2023RSE}.

Metal hydrides provide a promising alternative by achieving high volumetric and gravimetric densities under moderate conditions\cite{Allendorf2022NC, Felderhoff2007PCCP}. Light-element hydrides such as MgH$_{2}$ and CaH$_{2}$ offer high hydrogen-to-metal ratios (H/M = 2.0), but binary hydrides generally suffer from thermodynamic extremes: overly stable compounds require prohibitively high desorption temperatures, while unstable compounds absorb hydrogen only under impractically high pressures\cite{Allendorf2022NC, Felderhoff2007PCCP, Yang2007JPCC}. Transition-metal hydrides add complexity through magnetic effects, which can either stabilize or destabilize the system\cite{Allendorf2022NC, Felderhoff2007PCCP}. Consequently, most binary hydrides are unsuitable for practical, reversible storage\cite{Allendorf2022NC, Felderhoff2007PCCP, Yang2007JPCC}.

\begin{figure}[!h] 
\centering
\includegraphics[width=8.6cm]{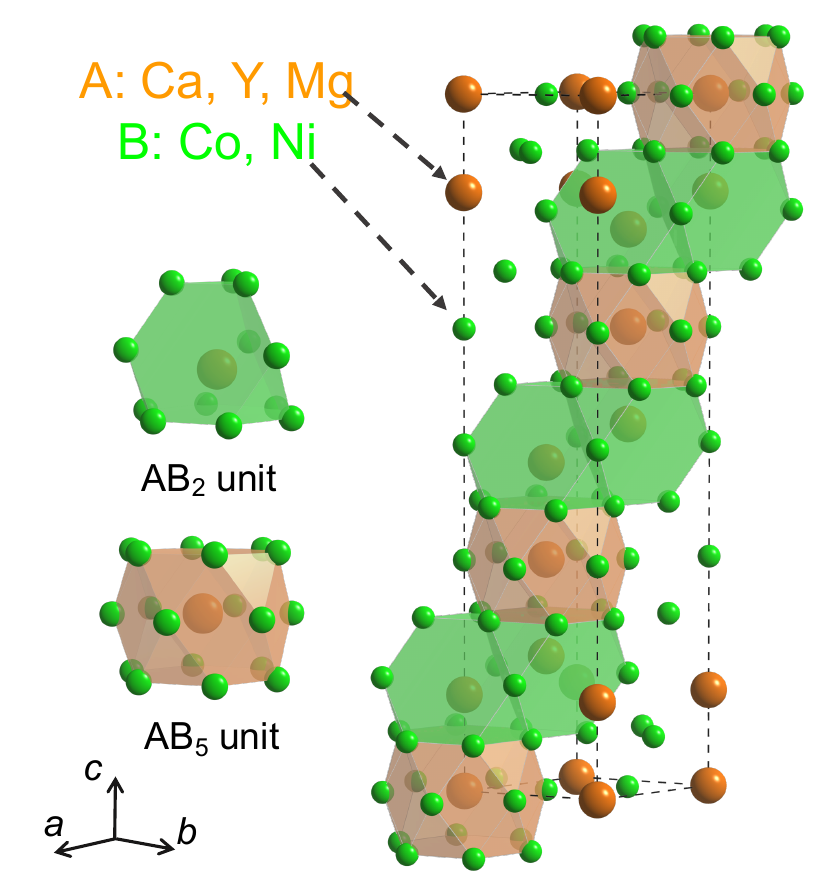} 
\caption{Crystal structure of stoichiometric AB$_{3}$ compounds (A = Ca, Y, Mg; B = Co, Ni) with rhombohedral space group $R\bar{3}m$ (No.~166). The structure can be conceptualized as a combination of AB$_{2}$ structural units (highlighted in orange) and AB$_{5}$ structural units (highlighted in green), representing two-thirds and one-third contributions, respectively. Large orange spheres represent A-site atoms (Ca, Y, or Mg) occupying both 3$a$ (AB$_{5}$ unit) and 6$c$ (AB$_{2}$ unit) Wyckoff positions, while smaller green spheres represent B-site atoms (Co or Ni).} 
\label{FIG1}
\end{figure}

Ternary intermetallic hydrides offer a balanced solution by combining intermediate thermodynamic stability with reversible hydrogen absorption and desorption\cite{Kadir1999JAC}. These systems are categorized into AB, AB$_{2}$, AB$_{3}$, A$_{2}$B$_{7}$, and AB$_{5}$ types, where A forms a stable hydride with strong hydrogen affinity and B forms an unstable hydride with low affinity\cite{Kadir1999JAC}. Among them, AB$_{3}$ alloys--including YCo$_{3}$, CaNi$_{3}$, and LaNi$_{3}$--crystallize in the rhombohedral $R\bar{3}m$ (PuNi$_{3}$-type) structure\cite{Kadir1999JAC, Pasquini2022PE}. This structure may be viewed as a hybrid of two-thirds of an AB$_{2}$ unit (ZrV$_{2}$ Laves phase) and one-third of an AB$_{5}$ unit (LaNi$_{5}$, CaCu$_{5}$ type)\cite{Pasquini2022PE, Iwase2025IJHE}, as illustrated in \textbf{Figure~\ref{FIG1}}. The theoretical H/M ratio of AB$_{3}$ hydrides reaches 1.75 (AB$_{3}$H$_{7}$), intermediate between AB$_{2}$ (H/M = 2.0) and AB$_{5}$ (H/M = 1.5)\cite{Pasquini2022PE, Iwase2025IJHE, Kadir1999JAC}. Compared to AB$_{5}$ alloys, AB$_{3}$ compounds provide higher gravimetric density, while unlike AB$_{2}$ systems, they combine moderate stability with favorable catalytic activity, enabling rapid hydrogen uptake\cite{Iwase2025IJHE, Kadir1999JAC}. Their rhombohedral structure also offers multiple interstitial sites that may host additional hydrogen under pressure\cite{Kadir1999JAC}. Furthermore, their magnetic properties can be tuned through composition, offering a route to optimize both stability and storage capacity\cite{Liu2003JAC}.

YCo$_{3}$ serves as a prototypical AB$_{3}$ alloy but suffers from limited gravimetric capacity due to the heavy yttrium atom\cite{Iwase2025IJHE, Liu2003JAC}. Its compositional modification is experimentally challenging, as stability and magnetism are highly sensitive to Co content\cite{Iwase2025IJHE, Liu2003JAC}. Sato \textit{et al.} demonstrated that partial substitution of Y with Mg in Y$_{x}$Mg$_{1-x}$Co$_{3}$ ($x = 0.58, 0.68$) enables reversible hydrogen cycling with a gravimetric density of $\sim$2.88 wt\% at room temperature under 10 GPa pressure\cite{Sato2025JPCC}. However, this composition approaches the intrinsic performance limit of Co-based systems due to strong spin-driven stabilization\cite{Sato2025JPCC}. By contrast, Ni-based AB$_{3}$ alloys such as Ca$_{x}$Y$_{y}$Mg$_{1-x-y}$Ni$_{3}$ achieve reversible hydrogen storage with high stability, suggesting that Ni substitution may overcome the magnetically constrained optimization bottleneck of Co-based systems\cite{Kadir1999JAC, Chen2000JAC, Zhang2005MSEB, Liang2003JAC, Sia2009IJHE, Kadir1999JAC2}. These advances underscore the urgent need for mechanistic insights into how magnetism controls thermodynamics and hydrogen storage performance.

In this work, we conduct a comprehensive analysis of stoichiometric AB$_{3}$ alloys (A = Ca, Y, Mg; B = Co, Ni) and their ternary solid solutions Ca$_{x}$Y$_{y}$Mg$_{1-x-y}$Co$_{3}$ and Ca$_{x}$Y$_{y}$Mg$_{1-x-y}$Ni$_{3}$. Chemical disorder is efficiently treated using the coherent potential approximation (CPA) within first-principles calculations, while formation energies are evaluated by averaging over the reference binaries (CaB$_{3}$, YB$_{3}$, and MgB$_{3}$) according to site fractions. Magnetic exchange interactions ($J_{ij}^{m}$) for the stoichiometric compounds are computed using the Liechtenstein linear-response formalism, and temperature-dependent magnetic properties are obtained through Monte Carlo simulations. This multiscale framework enables us to elucidate the fundamental coupling among magnetism, alloy stability, and hydrogen storage performance. In particular, we reveal that strong ferromagnetism in Co-based alloys destabilizes the AB$_{3}$ phase, while Ni substitution suppresses magnetism and restores thermodynamic stability, thereby resolving the intrinsic magnetism--stability trade-off that constrains optimization of hydrogen storage alloys.

\section{Methods}
The crystal structures of stoichiometric AB$_{3}$ compounds (A = Ca, Y, Mg; B = Co, Ni) with rhombohedral space group $R\bar{3}m$ (No.~166), as illustrated in \textbf{Figure~\ref{FIG1}}, were fully optimized using the \textit{Vienna Ab-initio Simulation Package} (VASP)\cite{VASP1, VASP2, VASP3}. The exchange--correlation functional was treated within the generalized gradient approximation (GGA) using the Perdew-Burke-Ernzerhof (PBE) functional\cite{PBE1996}. Structural optimizations employed a Monkhorst--Pack $k$-point mesh of $6 \times 6 \times 3$ for the rhombohedral unit cell and a plane-wave cutoff energy of 500 eV. Both lattice parameters and atomic positions were fully relaxed until residual forces on all atoms were less than 0.001 eV/$\angstrom$ and the energy convergence criterion was set to $10^{-6}$ eV. The formation energies of these stoichiometric AB$_{3}$ compounds were calculated using VASP and subsequently used to rescale the alloy formation energies obtained from CPA calculations (see \textbf{Supporting Information} for detailed lattice parameters and formation energy values).

The electronic structure and magnetic properties of Ca$_{x}$Y$_{y}$Mg$_{1-x-y}$Co$_{3}$ and Ca$_{x}$Y$_{y}$Mg$_{1-x-y}$Ni$_{3}$ alloys were investigated using the coherent potential approximation (CPA) as implemented in the SPR-KKR (Spin-Polarized Relativistic Korringa-Kohn-Rostoker) code\cite{SPRKKR1, SPRKKR2}. The CPA method enables efficient treatment of chemical disorder by replacing the random alloy with an effective medium, making it computationally tractable to explore the entire compositional space.

For the CPA calculations, three reference crystal structures of the stoichiometric compounds ACo$_{3}$ and ANi$_{3}$ (A = Ca, Y, Mg) were employed to compute alloy total energies across different local environments. The alloy formation energy was calculated as:

\begin{align} 
\Delta E_{\mathrm{form}}^{\mathrm{Ca}_{x}\mathrm{Y}_{y}\mathrm{Mg}_{1-x-y}\mathrm{B}_{3}} 
&= \frac{1}{4} \Big[ E\!\left(\mathrm{Ca}_{x}\mathrm{Y}_{y}\mathrm{Mg}_{1-x-y}\mathrm{B}_{3}\right) \notag \\ 
&\quad - x\,E(\mathrm{Ca}) - y\,E(\mathrm{Y}) - (1-x-y)\,E(\mathrm{Mg}) - 3\,E(\mathrm{B}) \Big], 
\label{Eq1} 
\end{align}

\noindent where $E$(Ca$_{x}$Y$_{y}$Mg$_{1-x-y}$B$_{3}$) is the total energy of the alloy per formula unit, and $E$(Ca), $E$(Y), $E$(Mg), and $E$(B) (B = Co, Ni) are the total energies per atom of the elemental constituents in their ground-state crystal structures.

Since CPA calculations are performed at fixed lattice parameters and cannot account for structural relaxation effects, we developed an averaging scheme to incorporate volume and local environment effects. Given the significant size differences among Ca, Y, and Mg atomic radii, these structural effects are expected to be substantial. The alloy formation energy was averaged over the three reference AB$_{3}$ structures according to the site fractions on the A sublattice:

\begin{align} 
\Delta E_{\mathrm{form}}^{\mathrm{Ca}_{x}\mathrm{Y}_{y}\mathrm{Mg}_{1-x-y}\mathrm{B}_{3}}(\text{average}) 
&= x \times \Delta E_{\mathrm{form}}^{\mathrm{Ca}_{x}\mathrm{Y}_{y}\mathrm{Mg}_{1-x-y}\mathrm{B}_{3}} (\text{CaB}_{3} \text{ structure}) \notag \\ 
&\quad + y \times \Delta E_{\mathrm{form}}^{\mathrm{Ca}_{x}\mathrm{Y}_{y}\mathrm{Mg}_{1-x-y}\mathrm{B}_{3}} (\text{YB}_{3} \text{ structure}) \notag \\ 
&\quad + (1-x-y) \times \Delta E_{\mathrm{form}}^{\mathrm{Ca}_{x}\mathrm{Y}_{y}\mathrm{Mg}_{1-x-y}\mathrm{B}_{3}} (\text{MgB}_{3} \text{ structure}). 
\label{Eq2} 
\end{align}

\noindent This weighted averaging scheme reflects the contributions of Ca, Y, and Mg to the alloy energetics based on their relative occupancies on the A site, providing a first-order approximation to account for volume and local structural effects that are not captured by the rigid-lattice CPA approach.

The lattice volumes of CaB$_{3}$, YB$_{3}$, and MgB$_{3}$ (B = Co, Ni) differ significantly, with MgB$_{3}$ being much smaller than CaB$_{3}$ and YB$_{3}$ due to small atomic radius of Mg. Although the coherent potential approximation (CPA) allows estimation of formation energies for (Ca, Y, Mg)B$_{3}$ alloys with disorder effects at moderate computational cost, it cannot optimize lattice parameters, limiting its ability to capture lattice volume-dependent effects. Our averaging scheme assumes that lattice volume changes primarily arise from the A-site element. As a result, the formation energy of A-rich corners (A: Ca, Y, Mg) approaches the formation energy of the corresponding stoichiometric AB$_{3}$ structure with fully relaxed by using VASP. This approach captures lattice volume-dependent effects more accurately than a simple equal-weight average, which will not approach the formation energy of the corresponding stoichiometric AB$_{3}$ structure.  Moreover, it includes the formation energies of two other structures, which are not the ground state structure of the corresponding stoichiometric composition. 

The theoretical limit of gravimetric hydrogen density (wt\%) of the metal hydride Ca$_{x}$Y$_{y}$Mg$_{1-x-y}$B$_{3}$H$_{7}$ (B = Co, Ni) is calculated using the molecular weight ratio of hydrogen to the total compound mass\cite{Sato2025JPCC}:

\begin{align} 
\text{wt\% of H$_{2}$} =
\frac{7\,M_{\mathrm{H}}}{
x M_{\mathrm{Ca}}
+ y M_{\mathrm{Y}}
+ (1-x-y) M_{\mathrm{Mg}}
+ 3 M_{\mathrm{B}}
+ 7 M_{\mathrm{H}}
}
\times 100\%
\label{Eq3} 
\end{align}

Magnetic exchange coupling constants ($J_{ij}^m$) for stoichiometric AB$_3$ alloys were calculated using the Liechtenstein formula within linear response theory as implemented in SPR-KKR\cite{Liechtenstein1987JMMM}. This approach provides a first-principles determination of the magnetic interactions by evaluating the response of the electronic system to small perturbations in the local magnetic moments. The corresponding classical Heisenberg Hamiltonian, including the effect of an external magnetic field, is expressed as\cite{HBT2022PRB, HBT2022AM}:

\begin{equation} 
H_{\mathrm{magnetic}} = -\sum_{\langle ij \rangle} J_{ij}^{m}\,\overrightarrow{S}_{i} \cdot \overrightarrow{S}_{j} - g\mu_{\mathrm{B}} \sum_{i} \overrightarrow{H}_{\mathrm{ext}} \cdot \overrightarrow{S}_{i}, 
\label{Eq4} 
\end{equation}

\noindent where the first term describes the exchange interactions between normalized spins $\overrightarrow{S}_{i}$ and $\overrightarrow{S}_{j}$, with $J_{ij}^{m} > 0$ favoring ferromagnetic (parallel) alignment and $J_{ij}^{m} < 0$ favoring antiferromagnetic (antiparallel) alignment. The second term represents the Zeeman interaction with an external magnetic field $\overrightarrow{H}_{\mathrm{ext}}$.

Temperature-dependent magnetic properties, including magnetization and magnetic phase transitions, were investigated using Monte Carlo simulations based on the Metropolis algorithm. The simulations employed a $16 \times 16 \times 8$ supercell containing 73,728 spins with periodic boundary conditions. Each Monte Carlo run consisted of 200,000 steps, with the first 100,000 steps discarded for thermalization to ensure equilibrium sampling. Magnetic susceptibility was calculated as the derivative of magnetization with respect to temperature, and Curie temperatures were determined from the peak positions in the susceptibility curves.

\section{Results and Discussion}

\subsection{ACo$_{3}$ Systems: Strong Magnetism Limits Stability}

\begin{figure}[!h] 
\centering
\includegraphics[width=16.8cm]{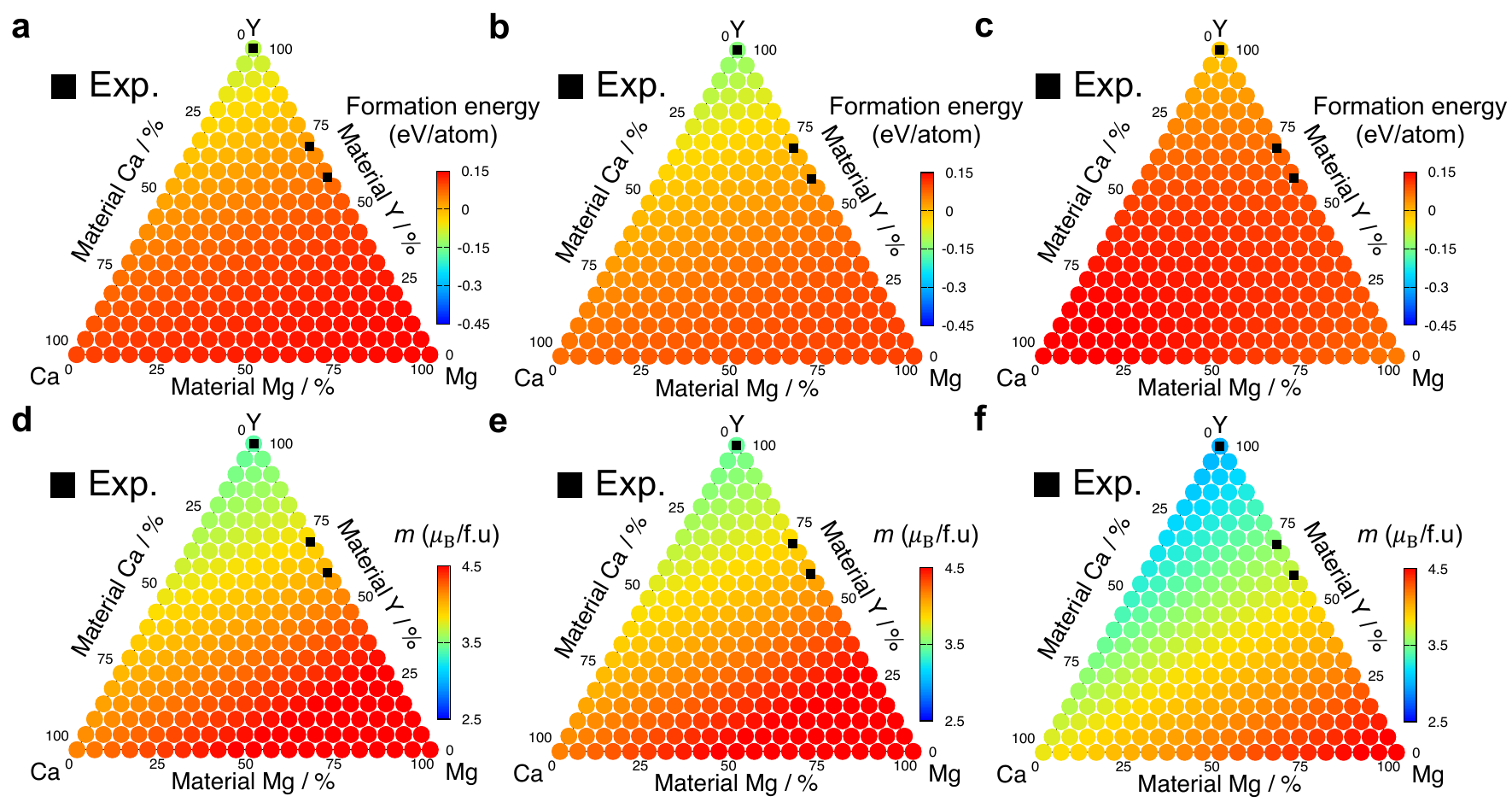} 
\caption{Formation energy diagrams of Ca$_{x}$Y$_{y}$Mg$_{1-x-y}$Co$_{3}$ calculated using Eq.~\ref{Eq1} with the crystal structures of CaCo$_{3}$ (a), YCo$_{3}$ (b), and MgCo$_{3}$ (c). Corresponding total magnetic moment diagrams of Ca$_{x}$Y$_{y}$Mg$_{1-x-y}$Co$_{3}$ with the crystal structures of CaCo$_{3}$ (d), YCo$_{3}$ (e), and MgCo$_{3}$ (f). Formation energies are given in eV per formula unit, and magnetic moments in $\mu_{\mathrm{B}}$ per formula unit. Black squares indicate compounds reported in previous experimental studies\cite{Liu2003JAC, Sato2025JPCC}.} 
\label{FIG2}
\end{figure}

The formation energy diagrams of Ca$_{x}$Y$_{y}$Mg$_{1-x-y}$Co$_{3}$ alloys, calculated using the lattice parameters of CaCo$_{3}$, YCo$_{3}$, and MgCo$_{3}$ as references, are shown in \textbf{Figure~\ref{FIG2}(a--c)}. Black squares indicate experimentally reported compositions\cite{Liu2003JAC, Sato2025JPCC}. Formation energies were evaluated using Eq.~\ref{Eq1} and subsequently averaged according to Eq.~\ref{Eq2}. Two primary factors govern the energy landscape: elemental composition and structural reference lattice. For each stoichiometric compound (CaCo$_{3}$, YCo$_{3}$, MgCo$_{3}$), the lowest formation energy occurs when adopting its own optimized lattice. Using CaCo$_{3}$ or YCo$_{3}$ lattice parameters, the minimum energies are found at the Y-rich corner, defining a thermodynamically stable triangular region (green to orange coloring) with negative formation energies. Within these structural frameworks, Ca-rich compositions show slightly higher energies, while Mg-rich compositions exhibit the largest values, indicating reduced stability. In contrast, when the MgCo$_{3}$ lattice is used as the reference, formation energies increase across most of the compositional range, with only the immediate Mg-rich region maintaining relatively low values. Nonetheless, the Y-rich corner still provides the most favorable energetics. This trend originates from the significantly smaller unit cell volume of MgCo$_{3}$ compared to CaCo$_{3}$ and YCo$_{3}$, which induces a strain effect when larger Ca or Y atoms are substituted (see \textbf{Supporting Information} for detailed lattice parameter comparisons). 

The corresponding total magnetic moment diagrams are presented in \textbf{Figure~\ref{FIG2}(d--f)}. A clear correlation emerges: the lowest total magnetic moments consistently occur at the Y-rich corner, whereas increasing Ca or Mg content enhances the magnetic moment. This trend parallels the formation energy distributions for the CaCo$_{3}$ and YCo$_{3}$ lattices, where the stable triangular Y-rich region corresponds to intermediate magnetic moments (green to yellow coloring). In contrast, the MgCo$_{3}$ lattice exhibits more complex magnetic behavior due to volume constraints, showing suppressed magnetization near the Y- and Ca-rich regions but overall larger variation in total magnetic moments.

\begin{figure*}[!h] 
\centering
\includegraphics[width=16.8cm]{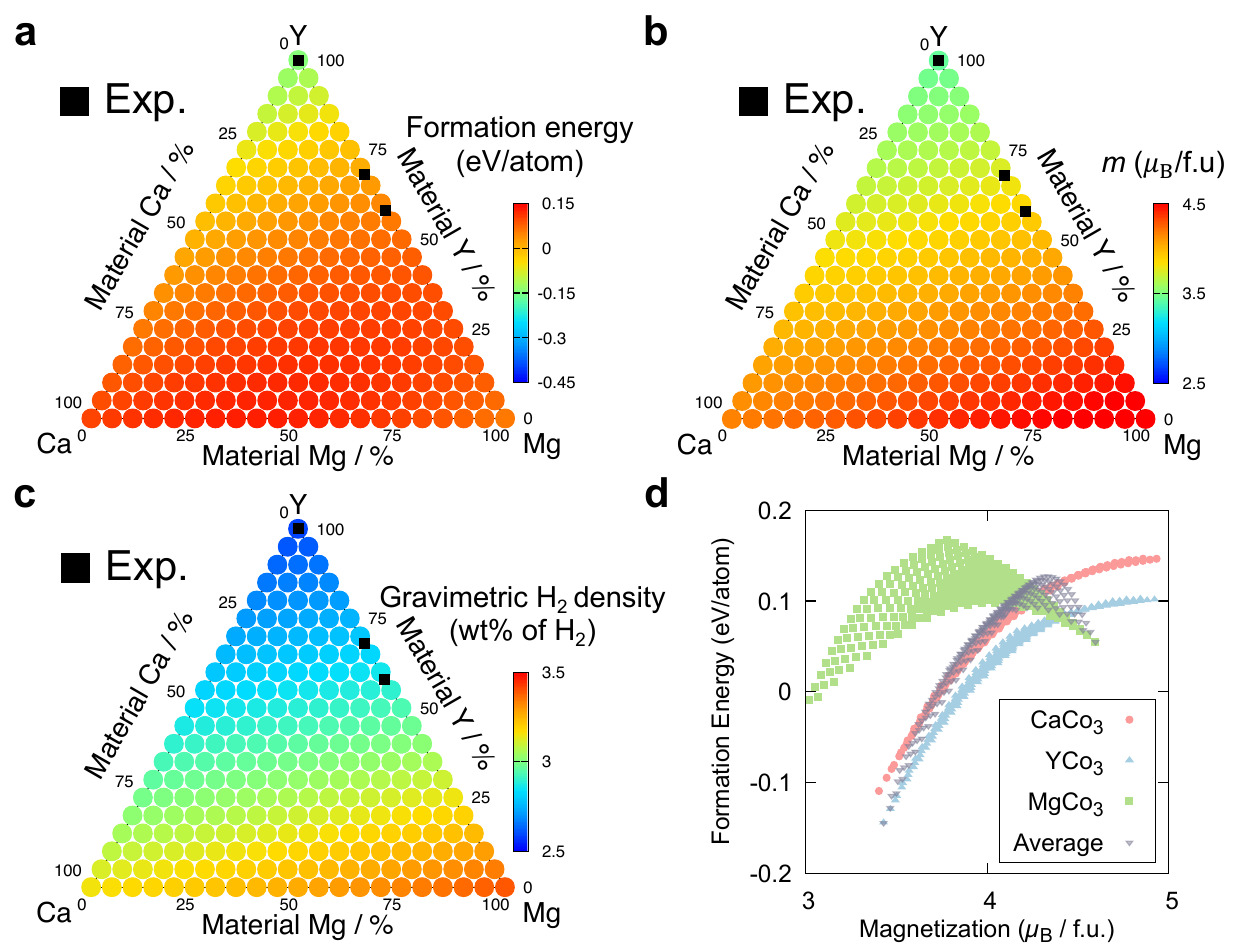} 
\caption{Average formation energy (a) and total magnetic moment (b) diagrams of Ca$_{x}$Y$_{y}$Mg$_{1-x-y}$Co$_{3}$ calculated using Eq.~\ref{Eq2}. (c) Theoretical gravimetric hydrogen density diagram of Ca$_{x}$Y$_{y}$Mg$_{1-x-y}$Co$_{3}$H$_{7}$ assuming complete hydrogenation from Eq.~\ref{Eq3}. Formation energies are given in eV per formula unit, magnetic moments in $\mu_{\mathrm{B}}$ per formula unit, and hydrogen densities in weight percent. Black squares indicate compounds reported in previous experimental studies\cite{Liu2003JAC, Sato2025JPCC}. (d) Formation energy versus total magnetic moment of Ca$_{x}$Y$_{y}$Mg$_{1-x-y}$Co$_{3}$ with CaCo$_{3}$ (light-red), YCo$_{3}$ (light-blue), and MgCo$_{3}$ (light-green) structures. The average formation energy versus average total magnetic moment (black) is estimated by using Eq.~\ref{Eq2}.} 
\label{FIG3}
\end{figure*}

The compositionally averaged formation energy and total magnetic moment diagrams of Ca$_{x}$Y$_{y}$Mg$_{1-x-y}$Co$_{3}$, obtained by weighted averaging over the three reference structures using Eq.~\ref{Eq2}, are shown in \textbf{Figure~\ref{FIG3}(a, b)}. The corresponding theoretical gravimetric hydrogen density of Ca$_{x}$Y$_{y}$Mg$_{1-x-y}$Co$_{3}$H$_{7}$ from Eq.~\ref{Eq3} is presented in \textbf{Figure~\ref{FIG3}(c)}. The averaged formation energy diagram reveals a stable triangular region concentrated near the Y-rich corner, where negative formation energies confirm thermodynamic feasibility. Within this region, Ca substitution is more favorable than Mg replacement, as the maximum Ca content for stability exceeds that of Mg. The color gradient from green to orange delineates this stability window, emphasizing that sufficient Y incorporation is essential for structural stability. This stability requirement, however, directly conflicts with hydrogen storage optimization. The gravimetric density map shows that higher Ca and Mg concentrations enhance hydrogen capacity due to their lower atomic masses relative to Y. Compositions at the Mg-rich corner reach the highest theoretical density ($\sim$3.39 wt\%), but these states possess positive formation energies and are thus unstable. The experimentally realized Y$_{0.68}$Mg$_{0.32}$Co$_{3}$ composition reported by Sato \textit{et al.} lies precisely at the edge of the stable triangular region, achieving a gravimetric density of $\sim$2.80 wt\%. This composition delivers performance comparable to the best Y$_{x}$Ca$_{1-x}$Co$_{3}$ alloys, yet highlights the intrinsic limitation of Co-based systems: further improvement is fundamentally constrained by the stability-capacity trade-off. Importantly, the stability map mirrors the magnetic moment distribution. The lowest total magnetic moments appear in the Y-rich corner and correlate with negative formation energies, whereas increasing Ca or Mg content enhances magnetization and destabilizes the lattice. These results establish that strong ferromagnetism is inherently detrimental to structural stability in Co-based AB$_{3}$ alloys.

The formation energy as a function of total magnetic moment for Ca$_{x}$Y$_{y}$Mg$_{1-x-y}$Co$_{3}$ alloys, evaluated using three reference structures and the averaged values from Eq.~\ref{Eq2}, is shown in \textbf{Figure~\ref{FIG3}(d)}. In the CaCo$_{3}$ and YCo$_{3}$ structures, which possess relatively large unit-cell volumes, a clear one-to-one correlation between total magnetic moment and formation energy is observed: the formation energy systematically increases with increasing magnetization. This monotonic behavior indicates that magnetism is the dominant factor governing alloy stability under these structural conditions. In contrast, the MgCo$_{3}$ structure, with its much smaller equilibrium volume, exhibits no such one-to-one correspondence--compositions with similar total magnetic moments show distinct formation energies. This breakdown of the correlation suggests that in the compact MgCo$_{3}$ lattice, magnetism couples more strongly with lattice volume and electronic degrees of freedom as the strain effect, thereby weakening the direct magnetism--stability dependence. In CaCo$_{3}$ and YCo$_{3}$, the lattice volumes are sufficiently large to sustain local magnetic moments, resulting in an almost linear dependence of formation energy on total magnetic moments. In MgCo$_{3}$, the smaller lattice volume introduces strain effects that perturb the relationship between formation energy and magnetization, leading to a non-linear behavior. This highlights the importance of lattice-volume effects on the coupling between magnetism and formation energy.

From the magnetization ternary diagrams in \textbf{Figure~\ref{FIG2}(d--f)}, compositions with low total magnetic moments (left-hand side) correspond to Y-rich regions near YCo$_{3}$, whereas those with high magnetic moments (right-hand side) correspond to Mg-rich regions near MgCo$_{3}$. Among the three end-member compounds, each structure attains its lowest formation energy when evaluated using its own lattice parameters. The average formation energy versus average total magnetic moment derived from Eq.~\ref{Eq2} smoothly approaches the corresponding stoichiometric structures as the A-site composition approaches their respective limits: starting from YCo$_{3}$ with the lowest formation energy, crossing the CaCo$_{3}$ structure, and finally reaching MgCo$_{3}$ with the highest magnetization.

\begin{table}[!h]
\caption{Magnetic moments of stoichiometric ACo$_{3}$ compounds (A = Ca, Y, Mg) from SPR-KKR calculations.}
\label{TAB1}
\centering
\begin{tabular}{|l|c|c|c|c|c|c|}
\hline
Materials & $m_{\text{total}}$ & $m_{\mathrm{A}(3a)}$ & $m_{\mathrm{A}(6c)}$ & $m_{\mathrm{Co}(3b)}$ & $m_{\mathrm{Co}(6c)}$ & $m_{\mathrm{Co}(18h)}$ \\
 ACo$_{3}$ & ($\mu_{\mathrm{B}}$/f.u.) & ($\mu_{\mathrm{B}}$/atom) & ($\mu_{\mathrm{B}}$/atom) & ($\mu_{\mathrm{B}}$/atom) & ($\mu_{\mathrm{B}}$/atom) & ($\mu_{\mathrm{B}}$/atom) \\
\hline
CaCo$_{3}$ & 4.155 & -0.254 & -0.320 & 1.532 & 1.637 & 1.426 \\
YCo$_{3}$  & 3.430 & -0.311 & -0.375 & 1.197 & 1.392 & 1.229 \\
MgCo$_{3}$ & 4.597 & -0.124 & -0.184 & 1.555 & 1.639 & 1.575 \\
\hline
\end{tabular}
\end{table}

\begin{figure*}[!h] 
\centering
\includegraphics[width=16.8cm]{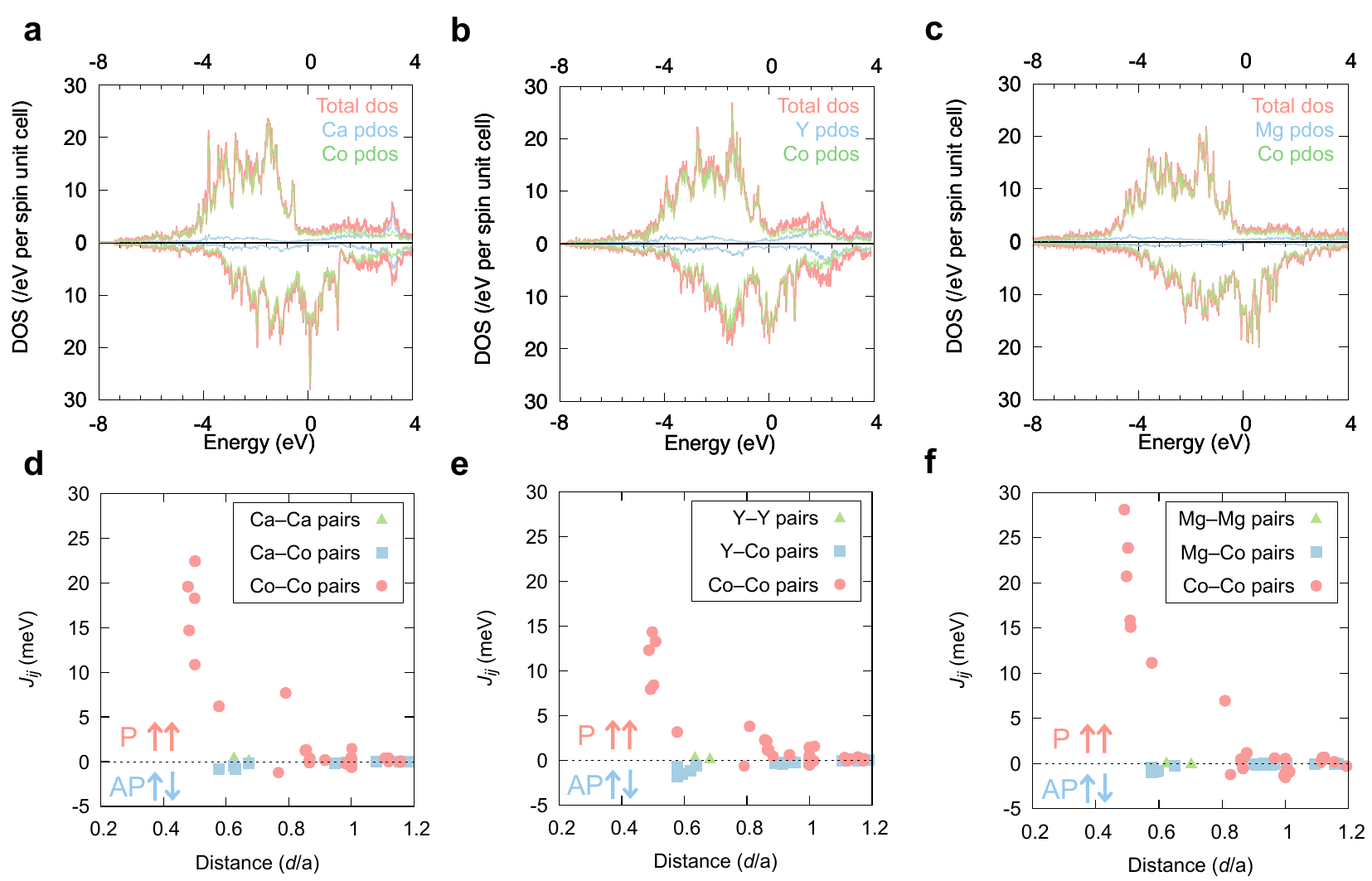} 
\caption{Total and partial density of states (DOS) of CaCo$_{3}$ (a), YCo$_{3}$ (b), and MgCo$_{3}$ (c). The total DOS is shown as light-red curves, while the partial DOS of A atoms (Ca, Y, Mg) and Co atoms are shown as light-blue and light-green curves, respectively. The partial DOS corresponds to the sum over all atoms of the same type in the unit cell. Magnetic exchange coupling constants of CaCo$_{3}$ (d), YCo$_{3}$ (e), and MgCo$_{3}$ (f), where A--A, A--Co, and Co--Co pairs are denoted by light-green triangles, light-blue squares, and light-red circles, respectively. Spin tends to parallel (P) in the case of $J_{ij} > 0$, while $J_{ij} < 0$ favors antiparallel (AP).} 
\label{FIG4}
\end{figure*}

To clarify the relationship between magnetism and stability, we analyze the electronic structures of stoichiometric CaCo$_{3}$, YCo$_{3}$, and MgCo$_{3}$. \textbf{Figure~\ref{FIG4}(a--c)} shows the total and partial densities of states (DOS). In all cases, the Co $d$ states dominate near the Fermi level, while the contributions from Ca, Y, and Mg are comparatively small. Magnetic moment analysis (\textbf{Table~\ref{TAB1}}) confirms that Co atoms carry the primary magnetic moments, whereas the A-site atoms (Ca, Y, Mg) contribute only small antiparallel moments, producing a ferrimagnetic configuration. A key distinction emerges in YCo$_{3}$, which exhibits a markedly reduced total magnetic moment (3.430 $\mu_{\mathrm{B}}$/f.u.) compared with CaCo$_{3}$ (4.155 $\mu_{\mathrm{B}}$/f.u.) and MgCo$_{3}$ (4.597 $\mu_{\mathrm{B}}$/f.u.). This reduction originates from Co atoms adopting a low-spin state due to weaker exchange splitting, combined with the larger antiferromagnetic contribution from Y. 

The magnetic exchange coupling constants in \textbf{Figure~\ref{FIG4}(d--f)} provide further insights. A--A interactions are negligible, while Co--Co couplings are strongly positive, confirming robust ferromagnetic interactions. A--Co couplings are weak and negative, consistent with the ferrimagnetic arrangement. Importantly, the Co--Co exchange couplings in YCo$_{3}$ are substantially weaker, reflecting its low-spin Co state. Consequently, the mean-field Curie temperature is significantly reduced to 754.6 K, compared with 1042.7 K for CaCo$_{3}$ and 1167.3 K for MgCo$_{3}$. These results demonstrate that Y substitution stabilizes the Co-based system by suppressing magnetic exchange through low-spin Co states, directly linking reduced magnetism to enhanced structural stability.

\begin{figure*}[!h] 
\centering
\includegraphics[width=8.0cm]{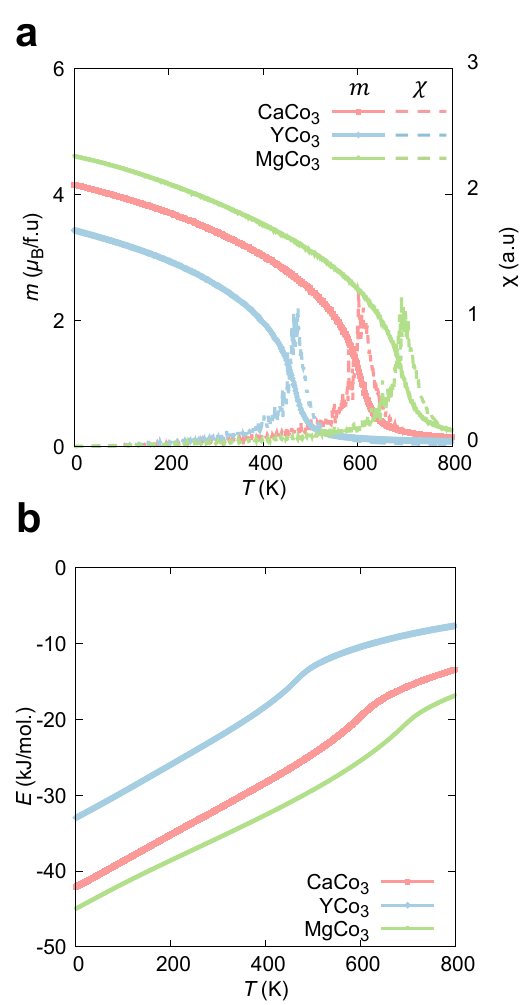} 
\caption{Temperature dependence of (a) magnetization and magnetic susceptibility, and (b) magnetic energy for CaCo$_{3}$, YCo$_{3}$, and MgCo$_{3}$. Magnetization and magnetic energy are shown as solid curves in light-red, light-blue, and light-green for CaCo$_{3}$, YCo$_{3}$, and MgCo$_{3}$, respectively. Magnetic susceptibility is shown as dashed curves with the same color scheme.} 
\label{FIG5}
\end{figure*}

Monte Carlo simulations yield more realistic estimates of the magnetic transition temperatures. As shown in \textbf{Figure~\ref{FIG5}(a)}, the temperature dependence of magnetization and susceptibility for CaCo$_{3}$, YCo$_{3}$, and MgCo$_{3}$ clearly reveals ferromagnetic-to-paramagnetic transitions, with Curie temperatures identified from the susceptibility maxima: 602 K (CaCo$_{3}$), 474 K (YCo$_{3}$), and 697 K (MgCo$_{3}$). These values, while lower than the mean-field estimates, preserve the same trend of suppressed ordering temperature in YCo$_{3}$. The Curie temperature of YCo$_{3}$ in previous experimental work is 301 K, which is lower than the value of Monte Carlo simulations\cite{Neznakhin2021JMMM}. The magnetic energy profiles in \textbf{Figure~\ref{FIG5}(b)} further highlight this behavior. All compounds exhibit large negative magnetic energies at 0 K, with MgCo$_{3}$ showing the most negative value, consistent with its highest saturation magnetization. The steepest energy variations occur at the Curie temperatures, reflecting the peak in magnetic heat capacity. Together, these results demonstrate that strong ferromagnetism, manifested in high magnetic moments and elevated Curie temperatures, correlates with reduced thermodynamic stability in Co-based AB$_{3}$ alloys. Notably, Mg incorporation, while favorable for achieving high hydrogen storage density, amplifies ferromagnetic interactions and thereby destabilizes the lattice. This intrinsic trade-off underscores the need to explore alternative B-site elements with weaker magnetic interactions to balance hydrogen storage capacity and structural stability.

\subsection{ANi$_{3}$ Systems: Reduced Magnetism Enables Enhanced Stability}

\begin{table}[!h]
\caption{Magnetic moments of stoichiometric ANi$_{3}$ (A = Ca, Y, Mg) from SPR-KKR.}
\label{TAB2}
\centering
\begin{tabular}{|l|c|c|c|c|c|c|}
\hline
Materials & $m_{\text{total}}$ & $m_{\mathrm{A}(3a)}$ & $m_{\mathrm{A}(6c)}$ & $m_{\mathrm{Ni}(3b)}$ & $m_{\mathrm{Ni}(6c)}$ & $m_{\mathrm{Ni}(18h)}$ \\
ANi$_{3}$ & ($\mu_{\mathrm{B}}$/f.u.) & ($\mu_{\mathrm{B}}$/atom) & ($\mu_{\mathrm{B}}$/atom) & ($\mu_{\mathrm{B}}$/atom) & ($\mu_{\mathrm{B}}$/atom) & ($\mu_{\mathrm{B}}$/atom) \\
\hline
CaNi$_{3}$ & 0.768 & $-0.058$ & $-0.070$ & 0.165 & 0.324 & 0.281 \\
YNi$_{3}$  & 0.000 & 0.000 & 0.000 & 0.000 & 0.000 & 0.000 \\
MgNi$_{3}$ & 0.793 & $-0.023$ & $-0.033$ & 0.198 & 0.312 & 0.274 \\
\hline
\end{tabular}
\end{table}

\begin{figure*}[!h] 
\centering
\includegraphics[width=16.8cm]{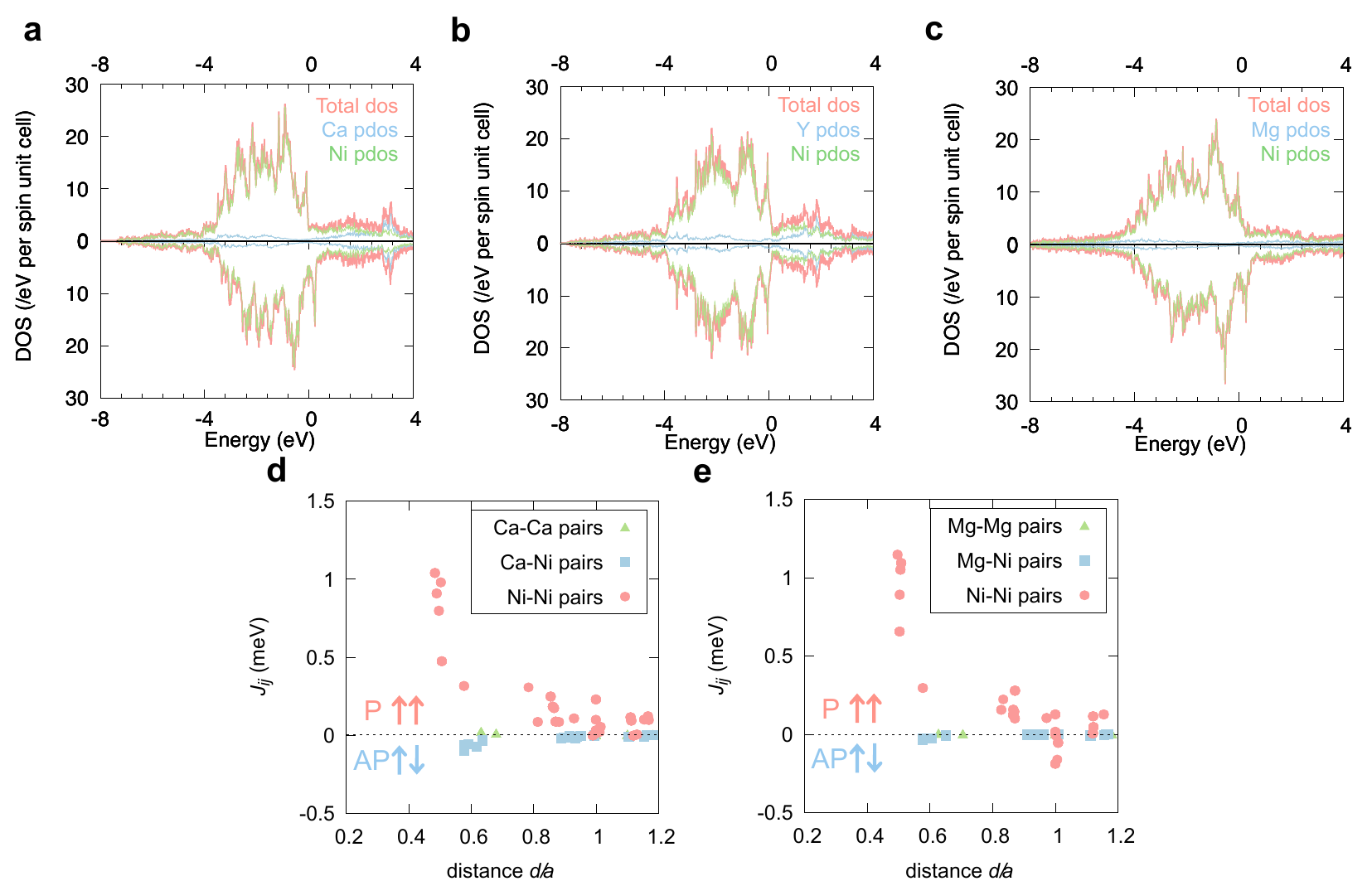} 
\caption{Total and partial density of states (DOS) of CaNi$_{3}$ (a), YNi$_{3}$ (b), and MgNi$_{3}$ (c). The total DOS is shown as light-red curves, while the partial DOS of A atoms (Ca, Y, Mg) and Ni atoms are shown as light-blue and light-green curves, respectively. The partial DOS corresponds to the sum over all atoms of the same type in the unit cell. Magnetic exchange coupling constants of CaNi$_{3}$ (d) and MgNi$_{3}$ (e), where A--A, A--Ni, and Ni--Ni pairs are denoted by light-green triangles, light-blue squares, and light-red circles, respectively. Spin tends to parallel (P) in the case of $J_{ij} > 0$, while $J_{ij} < 0$ favors antiparallel (AP).} 
\label{FIG6}
\end{figure*}

The electronic structures of CaNi$_{3}$, YNi$_{3}$, and MgNi$_{3}$, shown in \textbf{Figure~\ref{FIG6}(a--c)}, reveal fundamentally different magnetic behavior compared to their Co-based counterparts, reshaping the stability landscape for hydrogen storage applications. Ni $d$ states dominate near the Fermi level, as in Co systems, but the exchange splitting is substantially weaker due to the nearly filled $d^9$ configuration of Ni, in contrast to the $d^7$ configuration of Co. Magnetic moment analysis (\textbf{Table~\ref{TAB2}}) highlights this transformation: CaNi$_{3}$ and MgNi$_{3}$ exhibit very small total moments of 0.768 and 0.793 $\mu_{\mathrm{B}}$/f.u., respectively--over five times smaller than their Co analogues (4.155 and 4.597 $\mu_{\mathrm{B}}$/f.u.). Most notably, YNi$_{3}$ is fully nonmagnetic, with zero moments on all atomic sites, in stark contrast to YCo$_{3}$ (3.430 $\mu_{\mathrm{B}}$/f.u.). This suppression arises from Ni approaching a filled $d^{10}$ configuration, where crystal field effects favor a low-spin or nonmagnetic state. In weakly magnetic CaNi$_{3}$ and MgNi$_{3}$, Ni atoms at different crystallographic positions (3$b$, 6$c$, 18$h$) display small but varying moments (0.165--0.324 $\mu_{\mathrm{B}}$/atom), reflecting local environment differences within the rhombohedral structure. A-site atoms contribute minor antiparallel moments, as in the Co systems, but with greatly reduced magnitude. The ferrimagnetic configuration is preserved, yet the overall magnetization is substantially suppressed. This dramatic reduction in magnetic interactions directly relaxes stability constraints, providing a broader compositional space for designing thermodynamically robust, high-capacity Ni-based AB$_{3}$ hydrogen storage alloys.

\begin{figure*}[!h] 
\centering
\includegraphics[width=8.0cm]{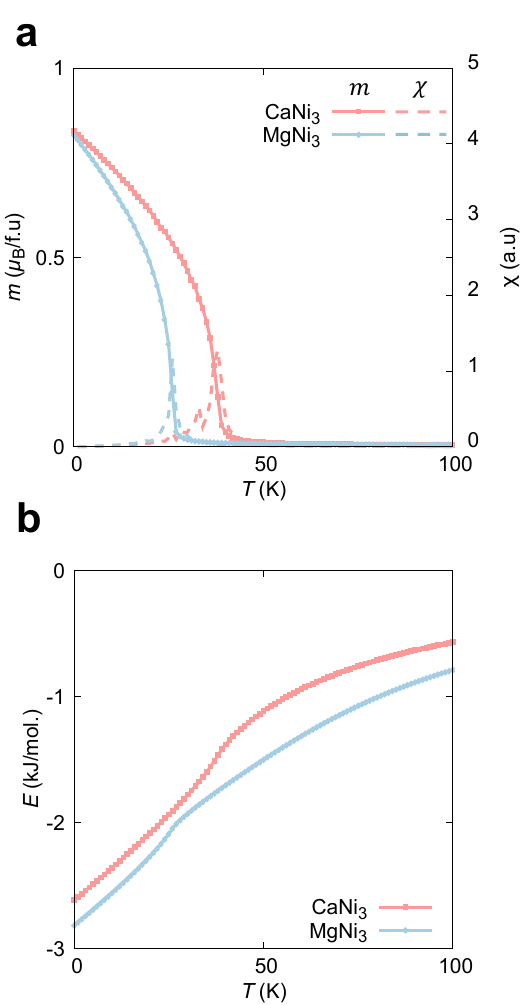} 
\caption{Temperature dependence of (a) magnetization and magnetic susceptibility, and (b) magnetic energy for CaNi$_{3}$ and MgNi$_{3}$. Magnetization and magnetic energy are shown as solid curves in light-red and light-blue for CaNi$_{3}$ and MgNi$_{3}$, respectively. Magnetic susceptibility is shown as dashed curves with the same color scheme.} 
\label{FIG7}
\end{figure*}

The magnetic exchange coupling constants for CaNi$_{3}$ and MgNi$_{3}$, shown in \textbf{Figure~\ref{FIG6}(d, e)}, are substantially weaker than those in Co-based systems. Ni--Ni interactions, while remaining ferromagnetic, are an order of magnitude smaller than the corresponding Co--Co couplings, directly reflecting the reduced magnetic moments on Ni sites and explaining the dramatically lower magnetic transition temperatures. A--Ni interactions remain weakly antiferromagnetic, consistent with the ferrimagnetic ground state, but their contribution to the overall magnetic behavior is minimal due to the weak Ni magnetism. Temperature-dependent magnetic properties, presented in \textbf{Figure~\ref{FIG7}}, further confirm this behavior. Mean-field estimates yield Curie temperatures of 88.6 K (CaNi$_{3}$) and 66.6 K (MgNi$_{3}$), substantially lower than Co analogues (1042.7 K and 1167.3 K, respectively). Monte Carlo simulations predict even lower transition temperatures of 38 K and 26 K, indicating that both compounds exhibit only weak ferromagnetism at very low temperatures and are effectively paramagnetic under typical hydrogen storage conditions. The magnetic energy evolution (\textbf{Figure~\ref{FIG7}(b)}) shows much smaller negative values at 0 K compared to Co-based systems, with steepest gradients near the Curie temperatures, reflecting the low magnetic stabilization energy. These results demonstrate that magnetic interactions in Ni-based AB$_{3}$ alloys play a minimal role in the total energy balance, fundamentally altering thermodynamic stability relative to strongly magnetic Co-based systems. Replacing Co with Ni transforms the electronic and magnetic landscape, removing the primary destabilizing factor in Co alloys and enabling access to previously inaccessible compositional regions. This reduction in magnetic interactions allows the design of thermodynamically stable alloys while preserving the structural framework necessary for high-capacity hydrogen storage.

The Liechtenstein formalism is generally suitable for estimating exchange parameters ($J_{ij}$) in itinerant magnets, such as Ni-based alloys. However, limitations arise from higher-order or non-Heisenberg interactions. For higher-order contributions, the magnetic force theorem remains valid but requires more complex multipolar rotations. Overall, these considerations provide a pathway to improve the accuracy of $J_{ij}$ while retaining the computational efficiency of the formalism.

\begin{figure*}[!h] 
\centering
\includegraphics[width=16.8cm]{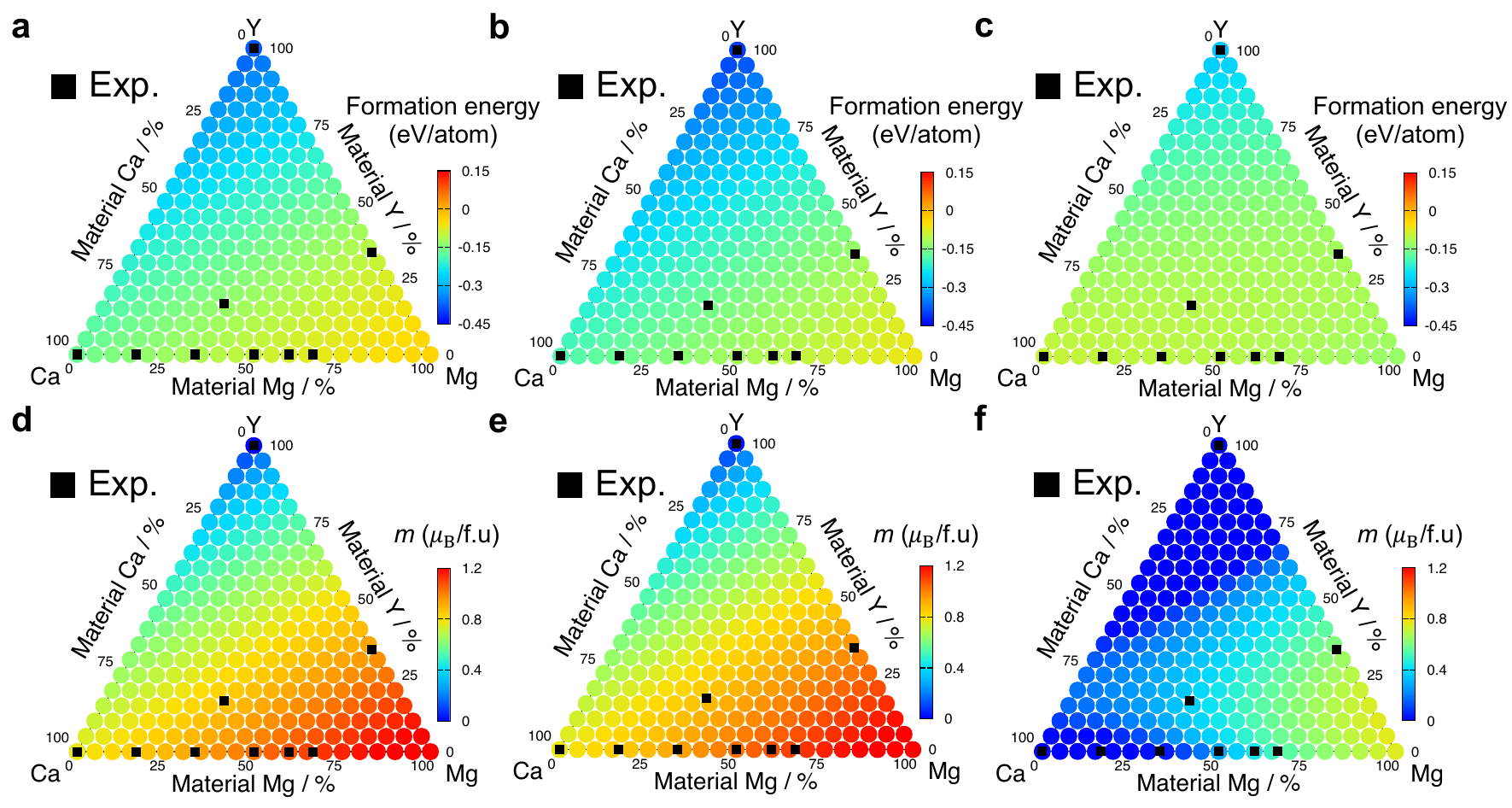} 
\caption{Formation energy diagrams of Ca$_{x}$Y$_{y}$Mg$_{1-x-y}$Ni$_{3}$ calculated using Eq.~\ref{Eq1} with the crystal structures of CaNi$_{3}$ (a), YNi$_{3}$ (b), and MgNi$_{3}$ (c). Corresponding total magnetic moment diagrams of Ca$_{x}$Y$_{y}$Mg$_{1-x-y}$Ni$_{3}$ with the crystal structures of CaNi$_{3}$ (d), YNi$_{3}$ (e), and MgNi$_{3}$ (f). Compounds reported in previous experimental studies are indicated by black squares\cite{Kadir1999JAC, Chen2000JAC, Zhang2005MSEB, Liang2003JAC, Sia2009IJHE, Kadir1999JAC2}.} 
\label{FIG8}
\end{figure*}

The formation energy diagrams of Ca$_{x}$Y$_{y}$Mg$_{1-x-y}$Ni$_{3}$ alloys, calculated using the three reference structures, are shown in \textbf{Figure~\ref{FIG8}(a--c)}. Compared to Co-based systems, a remarkable transformation is observed: all compositions across the ternary diagram exhibit negative formation energies, indicating universal thermodynamic stability. While the Y-rich corner remains the most stable region, the energy differences across the compositional space are substantially reduced, reflecting a more uniform stability landscape. The total magnetic moment diagrams in \textbf{Figure~\ref{FIG8}(d--f)} further illustrate this trend. YNi$_{3}$ retains zero magnetization across all reference structures, whereas compositions with increasing Ca and Mg content develop only weak ferromagnetism. Crucially, the magnetic moments remain far smaller than those in Co-based alloys, consistent with the enhanced thermodynamic stability. This analysis demonstrates that the suppression of magnetic interactions in Ni-based AB$_{3}$ alloys fundamentally alleviates the magnetism-driven stability constraints observed in Co systems, opening broad compositional space for stable, high-capacity hydrogen storage materials.

In YB$_{3}$ and CaB$_{3}$ (B = Co, Ni) crystal structures, the lattice volume is sufficient to sustain local magnetic moments on Co and Ni sites. This results in a near-linear dependence of formation energy on total magnetic moments, since the magnetic contribution to formation energy correlates directly with composition changes. In MgCo$_{3}$, however, the smaller lattice volume introduces additional strain effects, which complicate the dependence of formation energy on total magnetic moments. Despite this, local magnetic moments of Co persist even in Ca- or Y-rich regions, as shown in \textbf{Figure~\ref{FIG2}(f)}. In contrast, Ni exhibits weaker magnetism than Co. In the MgNi$_{3}$ structure, local magnetic moments of Ni can be quenched in Ca- or Y-rich regions due to small lattice volumes, leading to a much stronger lattice-volume effect compared with Co-based alloys, as shown in \textbf{Figure~\ref{FIG8}(f)}.

\begin{figure*}[!h] 
\centering
\includegraphics[width=16.8cm]{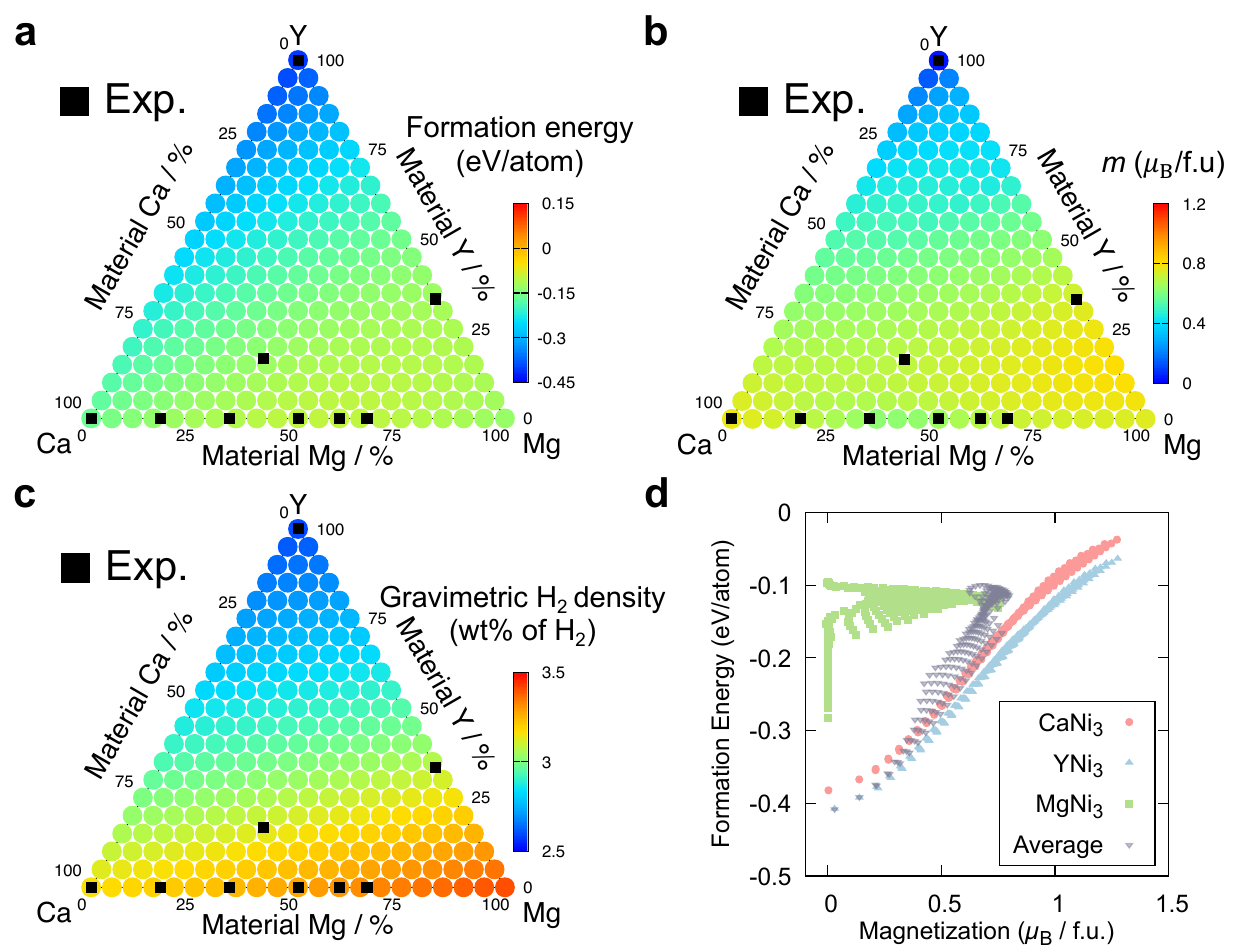} 
\caption{Average formation energy (a) and total magnetic moment (b) diagrams of Ca$_{x}$Y$_{y}$Mg$_{1-x-y}$Ni$_{3}$ calculated using Eq.~\ref{Eq2}. (c) Theoretical gravimetric hydrogen density of Ca$_{x}$Y$_{y}$Mg$_{1-x-y}$Ni$_{3}$H$_{7}$ (wt\%) from Eq.~\ref{Eq3}. Formation energies are given in eV per formula unit and magnetic moments in $\mu_{\mathrm{B}}$ per formula unit. Black squares indicate compounds reported in previous experimental studies\cite{Kadir1999JAC, Chen2000JAC, Zhang2005MSEB, Liang2003JAC, Sia2009IJHE, Kadir1999JAC2}. (d) Formation energy versus total magnetic moment of Ca$_{x}$Y$_{y}$Mg$_{1-x-y}$Ni$_{3}$ with CaNi$_{3}$ (light-red), YNi$_{3}$ (light-blue), and MgNi$_{3}$ (light-green) structures. The average formation energy versus average total magnetic moment (black) is estimated by using Eq.~\ref{Eq2}.} 
\label{FIG9}
\end{figure*}

The averaged formation energy and total magnetic moment diagrams for Ca$_{x}$Y$_{y}$Mg$_{1-x-y}$Ni$_{3}$ are presented in \textbf{Figure~\ref{FIG9}(a, b)}, with the corresponding theoretical gravimetric hydrogen densities from Eq.~\ref{Eq3} shown in \textbf{Figure~\ref{FIG9}(c)}. The universally negative formation energies mark a significant improvement over Co-based systems, as compositions with high hydrogen storage potential are now thermodynamically accessible. The hydrogen density distribution closely mirrors that of the Co system, with maximum values near the Mg-rich corner reaching $\sim$3.40 wt\% for pure MgNi$_{3}$H$_{7}$. Importantly, the experimentally known composition CaMg$_{2}$Ni$_{9}$ (Ca$_{0.33}$Mg$_{0.67}$Ni$_3$) achieves both high hydrogen storage density ($\sim$3.32 wt\%) and sufficient thermodynamic stability\cite{Kadir1999JAC, Kadir1999JAC2}. This indicates a moderate improvement in hydrogen storage capacity when using MgNi$_{3}$ compared with CaMg$_{2}$Ni$_{9}$. Most significantly, our calculations identify the previously unexplored Mg-rich corner of the ternary diagram as a promising compositional region that combines high theoretical hydrogen capacity with robust stability. This finding extends beyond known experimental compositions, highlighting new synthetic targets for high-performance hydrogen storage materials. The key enabling factor is the weak magnetism in Ni-based systems. Unlike Co-based alloys, where strong ferromagnetism imposes a fundamental stability barrier for high-capacity compositions, Ni-based AB$_{3}$ alloys maintain thermodynamic stability across the entire compositional range while preserving comparable hydrogen storage densities. This demonstrates that suppression of magnetic interactions is critical for reconciling high hydrogen capacity with structural stability in AB$_{3}$-type hydrides.

Although our calculations indicate that the Mg-rich region is thermodynamically stable, some experimental challenges may arise. Mg with high volatility requires precise stoichiometry control, and minor secondary phases may form if synthesis conditions are not optimized. Additionally, Mg-containing alloys may need careful surface treatment to ensure efficient hydrogen activation. These are typical experimental factors and do not represent fundamental limitations. Overall, the predicted stability and hydrogen storage capacity of Mg-rich Ni-based alloys remain robust, with only routine synthesis considerations expected. 

The formation energy as a function of total magnetic moment for Ca$_{x}$Y$_{y}$Mg$_{1-x-y}$Ni$_{3}$ alloys, evaluated using three reference structures and the averaged values from Eq.~\ref{Eq2}, is shown in \textbf{Figure~\ref{FIG9}(d)}. Unlike the Co-based alloys, the Ni-based alloys system exhibits considerably lower magnetic moments, ranging from zero to approximately 1.5 $\mu_{\text{B}}$/f.u., reflecting the weaker ferromagnetic character of Ni compared to Co. The reference structures CaNi$_{3}$ and YNi$_{3}$ display a similar trend to CaCo$_{3}$ and YCo$_{3}$, where the formation energy increases with increasing magnetization. This one-to-one correlation in the CaNi$_{3}$ and YNi$_{3}$ structures suggests that the Ni-based alloys behave similarly to the Co-based alloys when evaluated with the large lattice volumes characteristic of CaNi$_{3}$ and YNi$_{3}$. On the other hand, MgNi$_{3}$, with its much smaller lattice volume, shows a complex dependence where magnetism can disappear at both the Y-rich and Ca-rich corners, while the Mg-rich region retains the highest total magnetic moment. The averaged formation energy, derived from Eq.~\ref{Eq2}, provides a smooth interpolation between the three structures, with compositions approaching YNi$_{3}$ (low magnetization) being energetically most favorable, and those near the higher-magnetization regime corresponding to Ca-enriched and Mg-enriched compositions.

Since AB$_{3}$ alloys can be viewed as composed of two-thirds AB$_{2}$ and one-third AB$_{5}$ building blocks, our computational scheme is likely to perform well for alloys with similar crystal structures. However, challenges may arise when applying this scheme to materials containing rare-earth elements, such as SmCo$_{5}$ or LaNi$_{5}$. In these cases, strong correlation effects of localized $f$-electrons are difficult to capture accurately within standard DFT, which could affect the predicted formation energies and magnetic properties.

In (Ca, Y, Mg)Co$_{3}$ alloys, formation energy correlates with the total magnetic moment, which in turn reflects chemical composition. Mg-rich alloys exhibit high total magnetic moments and correspondingly higher formation energies (reduced stability), whereas Y-rich alloys suppress magnetism (low-spin state of Co and non-magnetic of Ni), which reduces the formation energy and enhances stability. Thus, magnetism and formation energy are strongly coupled, and the magnetic behavior, represented by the total magnetization, can serve as an indicator of alloy stability through its correlation with formation energy. The mechanism controlling the formation energy originates from the A-site composition (Ca, Y, or Mg), which strongly couples to the magnetic strength, such as the total magnetic moment.

On the other hand, the reason why the formation energy is reduced by replacing Co with Ni might come from several reasons, such as the electronic structure, $d$-band filling, and weaker magnetic energy penalty of Ni-based alloys. Ca, Y, and Mg donate electrons to form an intermetallic compound with Co and Ni. In the Co case, an unfilled d-band is less energetically favorable with charge transfer because more antibonding states will be occupied with high formation energy. Meanwhile, the nearly filled d-band of Ni becomes fully filled, which strengthens the bonding of the intermetallic compound, leading to lower formation energy. In the Co case, the material often pays an energy cost to maintain a high magnetic moment or spin polarization. Meanwhile, the Ni-based alloy has a smaller magnetic moment with a smaller energy penalty for spin polarization. Moreover, many Ni-based intermetallic alloys, such as LaNi$_{5}$, and CeNi$_{5}$ are very stable for hydrogen storage, while the Co-based intermetallic alloy is more rare due to high formation energy and magnetic instability.

\section{Summary}

Our comprehensive investigation establishes the decisive role of magnetism in governing the structural stability of AB$_{3}$-type alloys and clarifies its impact on hydrogen storage optimization. Across the Ca$_{x}$Y$_{y}$Mg$_{1-x-y}$Co$_{3}$ and Ca$_{x}$Y$_{y}$Mg$_{1-x-y}$Ni$_{3}$ systems, formation energy is strongly correlated with total magnetic moment. In Co-based alloys with large lattice volumes, such as CaCo$_{3}$ and YCo$_{3}$, the formation energy increases monotonically with magnetization, indicating that strong ferromagnetism intrinsically destabilizes the AB$_{3}$ lattice. This magnetism--stability correlation weakens in MgCo$_{3}$, where magnetic, electronic, and lattice degrees of freedom become strongly coupled. While Mg-rich regions in Co-based alloy offer high gravimetric densities, their stability is limited by strong magnetism, requiring Y substitution to suppress Co moments through a low-spin state--an intrinsic constraint to hydrogen storage optimization. In contrast, Ni substitution fundamentally reshapes this landscape: YNi$_{3}$ is nonmagnetic, and CaNi$_{3}$ and MgNi$_{3}$ exhibit only weak spin polarization, leading to thermodynamic stability across a wide composition range. The experimentally reported CaMg$_{2}$Ni$_{9}$ exemplifies this magnetic suppression strategy, combining high capacity ($\sim$3.32 wt\%) with good reversibility, while the Mg-rich Ni-based region identified in this work unites negative formation energies with the highest theoretical densities (up to $\sim$3.40 wt\%). Overall, these results establish magnetism as a key thermodynamic parameter and provide quantitative design principles for stabilizing high-capacity AB$_{3}$-type hydrogen storage materials through magnetic control. We also note that, this interesting phenomenon not only provides new design guidelines for hydrogen storage materials, but also for other types of materials and chemical reactions where AB3-type alloys can play an essential role (e.g., batteries and catalysis).

\section*{Author contributions}
H. B. Tran: Investigation, Writing - Original Draft. T. Sato: Writing - Review \& Editing. R. Sato: Writing - Review \& Editing. H. Saitoh: Writing - Review \& Editing. S. Orimo: Funding acquisition, Supervision, Writing - Review \& Editing. H. Li: Supervision, Resources, Writing - Review \& Editing.

\section*{Conflicts of interest}
There are no conflicts to declare.

\section*{Supporting Information}
Derivation of H/M expressions for AB$_{2}$, AB$_{3}$, and AB$_{5}$ alloys; rescaling formula for formation energies; lattice parameters for all stoichiometric AB$_{3}$ compositions (A = Ca, Y, Mg; B = Co, Ni).

\section*{Acknowledgements}
This work was supported by The Green Technologies of Excellence (GteX) Program, Japan (Grant No. JPMJGX23H1).

\section*{Data availability}
The data that support the findings of this study are available from the corresponding author upon reasonable request. Besides, the key data have also been uploaded to Digital Hydrogen Platform ($DigHyd$: \url{www.dighyd.org}).


\begin{thebibliography}{99}
\bibitem{Staffell2019EES} Staffell, I.; Scamman, D.; Abad, A. V.; Balcombe, P.; Dodds, P. E.; Ekins, P.; Shah, N.; Ward, K. R. The role of hydrogen and fuel cells in the global energy system. {\it Energy Environ. Sci.} {\bf 2019}, {\it 12}, 463--491. https://doi.org/10.1039/C8EE01157E.
\bibitem{Pollet2024IJHE} Boretti, A.; Pollet, B. G. Hydrogen economy: Paving the path to a sustainable, low-carbon future. {\it Int. J. Hydrog. Energy}, {\bf 2024}, {\it 93}, 307--319. https://doi.org/10.1016/j.ijhydene.2024.10.350.
\bibitem{Ahmad2024ACSSRM} Mehtab, A.; Ali, S. A.; Sadiq, I.; Shaheen, S.; Khan, H.; Fazil, M.; Pandit, N. A.; Naaz, F.; Ahmad, T. Hydrogen Energy as Sustainable Energy Resource for Carbon-Neutrality Realization. {\it ACS Sustain. Resource Manag.}, {\bf 2024} {\it 1}, 604--620. https://doi.org/10.1021/acssusresmgt.4c00039
\bibitem{Oliveira2021COCE} Oliveira, A. M.; Beswick, R. R.; Yan, Y. A green hydrogen economy for a renewable energy society. {\it Curr. Opin. Chem. Eng.}, {\bf 2021}, {\it 33}, 100701. https://doi.org/10.1016/j.coche.2021.100701.
\bibitem{Eriksson2017AE} Eriksson, E. L. V.; Gray, E. M. A. Optimization and integration of hybrid renewable energy hydrogen fuel cell energy systems--A critical review. {\it Appl. Energy}, {\bf 2017}, {\it 202}, 348--364. https://doi.org/10.1016/j.apenergy.2017.03.132.
\bibitem{Allendorf2022NC} Allendorf, M. D.; Stavila, V.; Snider, J. L.; Witman, M.; Bowden, M. E.; Brooks, K.; Tran, B. L.;  Autrey, T. Challenges to developing materials for the transport and storage of hydrogen. {\it Nat. Chem.}, {\bf 2022}, {\it 14}, 1214--1223. https://doi.org/10.1038/s41557-022-01056-2.
\bibitem{Felderhoff2007PCCP} Felderhoff, M.; Weidenthaler, C.; Helmoltb, R. V.; Eberleb, U. Hydrogen storage: the remaining scientific and technological challenges. {\it Phys. Chem. Chem. Phys.}, {\bf 2007} , {\it 9}, 2643--2653. https://doi.org/10.1039/B701563C. 
\bibitem{Deng2020J} Deng, J.; Bae, C.; Denlinger, A.; Miller, T. Electric Vehicles Batteries: Requirements and Challenges. {\it Joule}, {\bf 2020}. {\it 4}, 511--515. https://doi.org/10.1016/j.joule.2020.01.013.
\bibitem{Zhang2023RSE} Zhang, T. ; Uratani, J.; Huang, Y.; Xu, L.; Griffiths, S.; Ding, Y. Hydrogen liquefaction and storage: Recent progress and perspectives. {\it Renew. Sust. Energ.}, {\bf 2023}, {\it 176}, 113204. https://doi.org/10.1016/j.rser.2023.113204.
\bibitem{Yang2007JPCC} Yang, J.; Sudik, A.; Wolverton, C. Destabilizing LiBH$_{4}$ with a Metal (M = Mg, Al, Ti, V, Cr, or Sc) or Metal Hydride (MH$_{2}$ = MgH$_{2}$, TiH$_{2}$, or CaH$_{2}$). {\it J. Phys. Chem. C}, {\bf 2007}, {\it 111}, 19134--19140. https://doi.org/10.1021/jp076434z.
\bibitem{Kadir1999JAC} Kadir, K.; Kuriyama, N.; Sakai, T.; Uehara, I.; Eriksson, L. Structural investigation and hydrogen capacity of CaMg$_{2}$Ni$_{9}$: a new phase in the AB$_{2}$C$_{9}$ system isostructural with LaMg$_{2}$Ni$_{9}$. {\it J. Alloy. Compd.}, {\bf 1999}, {\it 284}, 145--154. https://doi.org/10.1016/S0925-8388(98)00965-7.
\bibitem{Pasquini2022PE} Pasquini, L.; Sakaki, K.; Akiba, E. et al. Magnesium- and intermetallic alloys-based hydrides for energy storage: modelling, synthesis and properties. {\it Prog. Energy}, {\bf 2022}, {\it 4}, 032007. https://doi.org/10.1088/2516-1083/ac7190.
\bibitem{Iwase2025IJHE} Iwase, K.; Kimura, T.; Mori, K. Y-Co binary alloys with superlattice structures: Relationship between hydrogenation cyclic stability and crystal structures of hydride phases. {\it Int. J. Hydrogen Energy}, {\bf 2025}, {\it 139}, 62--68. https://doi.org/10.1016/j.ijhydene.2025.05.258.
\bibitem{Liu2003JAC} Liu, J.; Broom, D. P.; Georgiev, P. A. L.; Ross, D. K. Magnetic properties of the YCo$_{3}$-H system. {\it J. Alloy. Compd.}, {\bf 2003}, {\it 356-357}, 174-177. https://doi.org/10.1016/S0925-8388(03)00260-3
\bibitem{Sato2025JPCC} Sato, T.; Saitoh, H.; Utsumi, R.; Ito, J.; Obana, K.; Nakahira, Y.; Sheptyakov, D.; Honda, T.; Sagayama, H.; Takagi, S.; Kono, T.; Yang, H.; Luo, W.; Lombardo, L.; Z\"uttel,  A.; Orimo, S. -i. Synthesis, crystal structure, and hydrogen storage properties of an ab3-based alloy synthesized by disproportionation reactions of AB$_{2}$-based alloys. {\it J. Phys. Chem. C}, {\bf 2025}, {\it 129}, 2865--2873. https://doi.org/10.1021/acs.jpcc.4c06759
\bibitem{Chen2000JAC} Chen, J.; Takeshita, H. T.; Tanaka, H.; Kuriyama, N.; Sakai, T.; Uehara, I.; Haruta, M. Hydriding properties of LaNi$_{3}$ and CaNi$_{3}$ and their substitutes with PuNi$_{3}$-type structure. {\it J. Alloy. Compd.}, {\bf 2000}, {\it 302}, 304--313. https://doi.org/10.1016/S0925-8388(00)00694-0. 
\bibitem{Zhang2005MSEB} Zhang, X.; Yin, W.; Chai, Y.; Zhao, M. Structure and electrochemical characteristics of RENi$_{3}$ alloy. {\it Mater. Sci. Eng. B}, {\bf 2005}, {\it 117}, 123-128. https://doi.org/10.1016/j.mseb.2004.11.017.
\bibitem{Liang2003JAC} Liang, G.; Schulz, R. Phase structures and hydrogen storage properties of Ca-Mg-Ni alloys prepared by mechanical alloying. {\it J. Alloy. Compd.}, {\bf 2003}, {\it 356-357}, 612-616. https://doi.org/10.1016/S0925-8388(02)01285-9.
\bibitem{Kadir1999JAC2} Kadir, K.;  Sakai, T.; Uehara, I. Structural investigation and hydrogen capacity of YMg$_{2}$Ni$_{9}$ and (Y$_{0.5}$Ca$_{0.5}$)(MgCa)Ni$_{9}$ : new phases in the AB$_{2}$C$_{9}$ system isostructural with LaMg$_{2}$Ni$_{9}$. {\it J. Alloy. Compd.}, {\bf 1999}, {\it 287}, 264--270. https://doi.org/10.1016/S0925-8388(99)00041-9. 
\bibitem{Sia2009IJHE} Sia, T. Z.; Zhang, Q. A.; Pang, G.; Liu, D. M.; Liu, N. Structural characteristics and hydrogen storage properties of Ca$_{3.0-x}$Mg$_{x}$Ni$_{9}$ (x=0.5, 1.0, 1.5 and 2.0) alloys. {\it Int. J. Hydrogen Energy}, {\bf 2009}, {\it 34}, 1483--1488. https://doi.org/10.1016/j.ijhydene.2008.11.051.
\bibitem{VASP1} Kresse, G.; Hafner, J. Ab Initio Molecular Dynamics for Liquid Metals. {\it Phys. Rev. B} {\bf 1993}, {\it 47}, 558. https://doi.org/10.1103/PhysRevB.47.558.
\bibitem{VASP2} Kresse, G.; Furthm\"uller, J. Efficiency of Ab-Initio Total Energy Calculations for Metals and Semiconductors Using a Plane-Wave Basis Set. {\it Comput. Mater. Sci.} {\bf 1996}, {\it 6}, 15--50. https://doi.org/10.1016/0927-0256(96)00008-0.
\bibitem{VASP3} Kresse, G.; Furthm\"uller, J. Efficient Iterative Schemes for Ab-Initio Total-Energy Calculations Using a Plane-Wave Basis Set. {\it Phys. Rev. B} {\bf 1996}, {\it 54}, 11169. https://doi.org/10.1103/PhysRevB.54.11169.
\bibitem{PBE1996} Perdew, J. P.; Burke, K.; Ernzerhof, M. Generalized Gradient Approximation Made Simple. {\it Phys. Rev. Lett.} {\bf 1996}, {\it 77}, 3865--3868. https://doi.org/10.1103/PhysRevLett.77.3865.
\bibitem{SPRKKR1} Ebert, H.; et al. The Munich SPR-KKR Package, version 8.6. Available at: https://www.ebert.cup.uni-muenchen.de/sprkkr8/licenses (Accessed 31 December 2022).
\bibitem{SPRKKR2} Ebert, H.; K\"odderitzsch, D.; Min\'ar, J. Calculating Condensed Matter Properties Using the KKR-Green's Function Method-Recent Developments and Applications. {\it Rep. Prog. Phys.} {\bf 2011}, {\it 74}, 096501. https://doi.org/10.1088/0034-4885/74/9/096501.
\bibitem{Liechtenstein1987JMMM} Liechtenstein, A. I.; Katsnelson, M. I.; Antropov, V. P. et al., Local spin density functional approach to the theory of exchange interactions in ferromagnetic metals and alloys. {\it{J. Magn. Magn. Mater.}} {\bf{1987}}, 67, 65--74. https://doi.org/10.1016/0304-8853(87)90721-9.
\bibitem{HBT2022PRB} Tran, H. B.; Momida, H.; Matsushita, Y.; Sato, K.; Makino, Y.; Shirai, K.; Oguchi, T. Effect of Magnetocrystalline Anisotropy on Magnetocaloric Properties of an AlFe$_{2}$B$_{2}$ Compound. {\it Phys. Rev. B} {\bf 2022}, {\it 105}, 134402. https://doi.org/10.1103/PhysRevB.105.134402.
\bibitem{HBT2022AM} Tran, H. B.; Momida, H.; Matsushita, Y.; Shirai, K.; Oguchi, T. Insight into Anisotropic Magnetocaloric Effect of CrI$_{3}$. {\it Acta Mater.} {\bf 2022}, {\it 231}, 117851. https://doi.org/10.1016/j.actamat.2022.117851.
\bibitem{Neznakhin2021JMMM} Neznakhin, D. S.; Bartashevich, A. M.; Volegov, A. S.; Bartashevich, M. I.; Andreev, A. V. Magnetic anisotropy in RCo$_{3}$ (R = Lu and Y) single crystals. {\it J. Magn. Magn. Mater.} {\bf 2021}, {\it 539}, 168367. https://doi.org/10.1016/j.jmmm.2021.168367.



\end{thebibliography}

\begin{figure*}[!h]
\centering
\includegraphics[width=16cm]{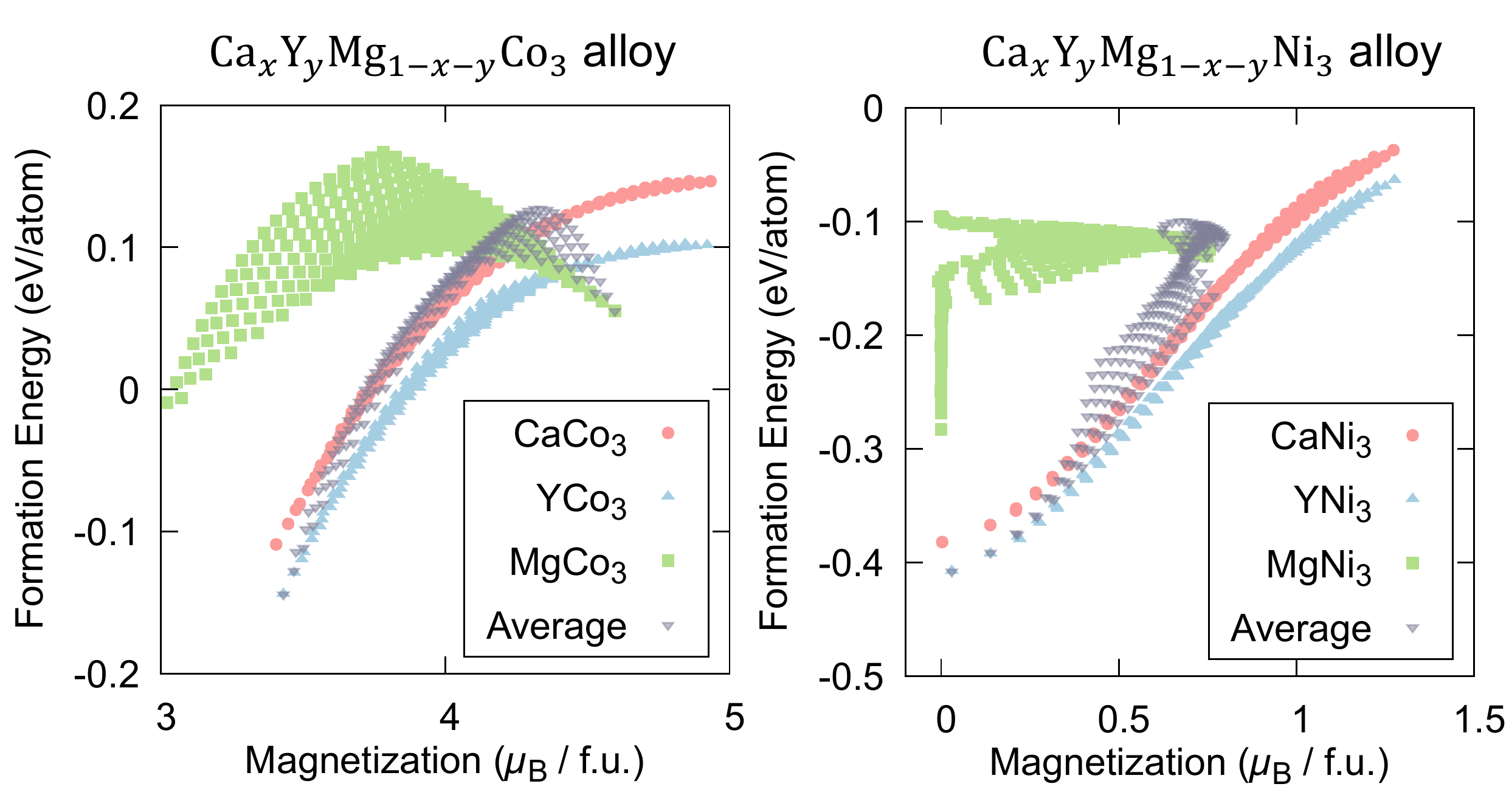}
\caption*{TOC}
\label{TOC}
\end{figure*}

\end{document}